\documentclass[final,5p,times,twocolumn]{elsarticle}

\usepackage{amssymb}

\journal{Nuclear Physics B}

\begin{document}

\begin{frontmatter}



\title {\textbf{Electric dipole strength and dipole polarizability in $^{48}$Ca within a fully self--consistent second random--phase approximation}}
\author{D. Gambacurta}
\address{Extreme Light Infrastructure - Nuclear Physics (ELI-NP), Horia Hulubei National Institute for Physics and Nuclear
Engineering, 30 Reactorului Street, RO-077125 M˘agurele, Jud. Ilfov, Romania}
\author{M. Grasso}
\address{Institut de Physique Nucl\'eaire, CNRS-IN2P3, Universit\'e Paris-Sud,
Universit\'e Paris-Saclay, 91406 Orsay, France}
\author{O. Vasseur}
\address{Institut de Physique Nucl\'eaire, CNRS-IN2P3, Universit\'e Paris-Sud,
Universit\'e Paris-Saclay, 91406 Orsay, France}

\begin{abstract}
The second random--phase--approximation model corrected by a subtraction 
procedure designed to cure double counting, instabilities, and ultraviolet 
divergences, is employed for the first time to analyze the dipole strength and 
polarizability in $^{48}$Ca. All the terms of the residual interaction are 
included, leading to a fully self-consistent scheme. Results are illustrated with 
two Skyrme parametrizations, SGII and SLy4. Those obtained with the SGII 
interaction are particularly satisfactory. In this case, the low--lying strength 
below the neutron threshold is extremely well reproduced and the giant dipole 
resonance is described in a very satisfactory way especially in its spreading 
and fragmentation. Spreading and fragmentation are produced in a natural way 
within such a theoretical model by the coupling of 1 particle--1 hole and 2 
particle--2 hole configurations. Owing to this feature, we may provide for 
the electric polarizability as a function of the excitation energy a curve with 
a similar slope around the centroid energy of the giant resonance compared to 
the corresponding experimental results. This represents a considerable 
improvement with respect to previous theoretical predictions obtained with the 
random--phase approximation or with several {\it ab--initio} models. In such 
cases, the spreading width of the excitation cannot be reproduced and the 
polarizability as a function of the excitation energy displays a stiff increase 
around the predicted centroid energy of the giant resonance. 
\end{abstract}


\end{frontmatter}


\section{Introduction}
Accurate measurements and theoretical computations of the electric dipole 
polarizability have strong implications in constraining the symmetry energy as 
well as its density dependence and slope \cite{roca}. These quantities are key 
ingredients in nuclear structure for describing  for example the neutron--skin 
thickness of neutron--rich nuclei \cite{warda, centelles} or in nuclear 
astrophysics for predicting the radius and the proton fraction of neutron stars  
as well as the neutron star cooling \cite{lp,lattimer,baldo}. Several experimental 
studies \cite{polosaka,osaka} aimed to obtain the full electric dipole response, and therefore the dipole polarizability,
have been  carried out in the last years at the RCNP facility in Osaka and systematic future studies
are planned at ELI-NP in Bucharest \cite{TDR}.

Recently, the electric dipole polarizability of $^{48}$Ca was determined at RCNP, Osaka, employing the ($p,p'$) reaction 
at forward angle \cite{osaka}. 
A comparison of the 
deduced dipole polarizability with the polarizability provided by {\it ab initio} calculations based on chiral effective--field--theory 
interactions and by energy--density--functional (EDF) models showed a reasonable agreement with some of these theoretical predictions, 
consistent with a neutron skin of $0.14-0.20$ fm in $^{48}$Ca. 

Less recently, the dipole response was measured below the 
neutron threshold ($\sim 10$ MeV) in $^{48}$Ca with the ($\gamma,\gamma'$) reaction \cite{low}. It was shown that 
only beyond mean--field models including correlations, such as 
the second random--phase approximation (SRPA) \cite{gamba1,gamba2,gambapygmy} based on the mixing between 1 particle-1 hole (1p1h) and 2 particle-2 hole (2p2h) configurations or 
the extended theory of finite Fermi systems (ETFFS) \cite{litvi1,litvi2}, based on the quasiparticle--phonon coupling,    
could account for this low--lying strength in $^{48}$Ca. 
It is worth noticing for example that random--phase approximation (RPA) predictions fail in reproducing such a low--lying strength either because the lowest RPA energies are larger than 10 MeV or (in those cases where a peak is found below 10 MeV) because the RPA model cannot provide any fragmentation in the strength distribution apart from the single--particle Landau damping. On the other hand, it is well known that the SRPA model is perfectly tailored to account for a fragmented strength owing to the presence of 2p2h configurations. A fragmented SRPA dipole strength below 10 MeV was indeed illustrated in Ref. \cite{gambapygmy} for $^{48}$Ca. However, several points remained open in that study. First, at that time, all SRPA 
results suffered essentially from two drawbacks: a too strong downwards shift of the spectrum with respect to the RPA excitation energies and an ultraviolet divergence (in cases where zero--range forces were used). 
The first of these drawbacks was eventually related to a more general stability problem (violation of the Thouless theorem, that can lead to imaginary solutions) \cite{papa}. In the case of Skyrme-- and 
Gogny--based SRPA models, a solution to both drawbacks was recently proposed and applied \cite{gamba2015,epja}, based on a subtraction procedure \cite{tse1,tse2}. 
As was discussed in Refs. \cite{tse1,tse2}, such a subtraction procedure was originally designed to remove the double counting of 
correlations in those cases where beyond mean--field calculations are performed employing effective interactions adjusted at the mean--field level. The risk of double counting is then avoided in the SRPA model implemented with the subtraction procedure in cases where traditional EDF functionals are used, based for instance on Skyrme or Gogny interactions. 
Another point that remained open in the study of Ref. \cite{gambapygmy} was that the SRPA $B(E1)$ transition probability, integrated up to 10 MeV, was found to be definitely much larger than the experimental value.  
We will show here that, by using the implemented SRPA model based on the subtraction method, we also remove the problem of  overestimation of the transition probability: some states having strong 1p1h components, which are too much shifted downwards by the standard SRPA, are pushed to higher energies by the subtraction, reducing in this way the $B(E1)$ value below 
10 MeV.     

In this work, we predict the dipole response of $^{48}$Ca in both the low--lying and the giant dipole resonance (GDR) regions of the excitation spectrum 
using the subtracted second random--phase approximation (SSRPA) model of Ref. \cite{gamba2015}. By comparing our results with the ($\gamma,\gamma'$) data for the low--lying part of the spectrum and with the 
recent ($p,p'$) data in the GDR region, we show the importance of including correlations in the SSRPA model for: (i) correctly describing the 
low--energy response, (ii) accounting for the spreading width in the GDR region, (iii) reproducing the increase of the electric dipole polarizability as a function of the excitation energy in the region of the GDR.  


\section{Dipole spectra and polarizability in $^{48}$Ca}
The formalism and implementation details of the SSRPA model can be found in Ref. \cite{gamba2015}. Previous
applications of the SRPA and SSRPA were done without including the spin-orbit and Coulomb terms in the residual interaction.
In this work, we apply for the first time a fully self--consistent scheme, including all the terms of the interaction
consistently with the mean--field Hartree--Fock description. This allows us to separate in a clean way the
physical spectrum from the spurious components of the center of mass motion appearing in the dipole channel.
Due to the inclusion of all terms of the residual interaction,
the present SRPA results are  different compared to those illustrated in Ref. 
\cite{gambapygmy}, where the spin--orbit and the Coulomb contributions were 
not taken into account. Two Skyrme parametrizations are employed in the present work, SGII 
\cite{sgii}, which was already employed in our previous study of Ref. 
\cite{gambapygmy} and SLy4 \cite{sly4}. 
\begin{figure}
\includegraphics[width=0.4\paperwidth]{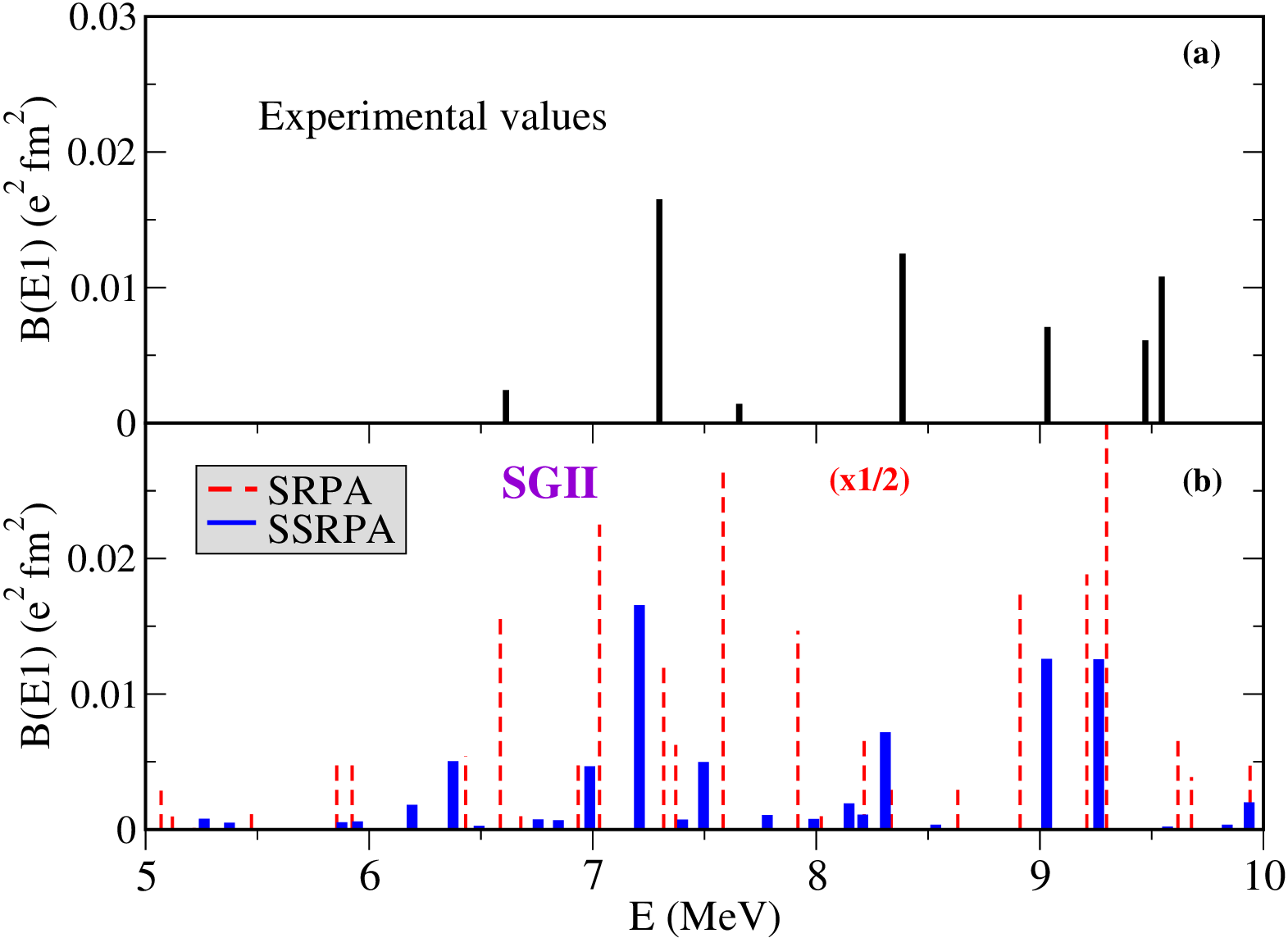}
\caption{(a) Experimental $B(E1)$ values \cite{low}; (b) Theoretical predictions for the transision probabilities $B(E1)$ calculated with the standard SRPA (dashed red bars; the values have been devided by 2) and with the SSRPA (blue thick bars), employing the Skyrme parametrization SGII.}
\label{fig1}
\end{figure}

\begin{figure}
\includegraphics[width=0.4\paperwidth]{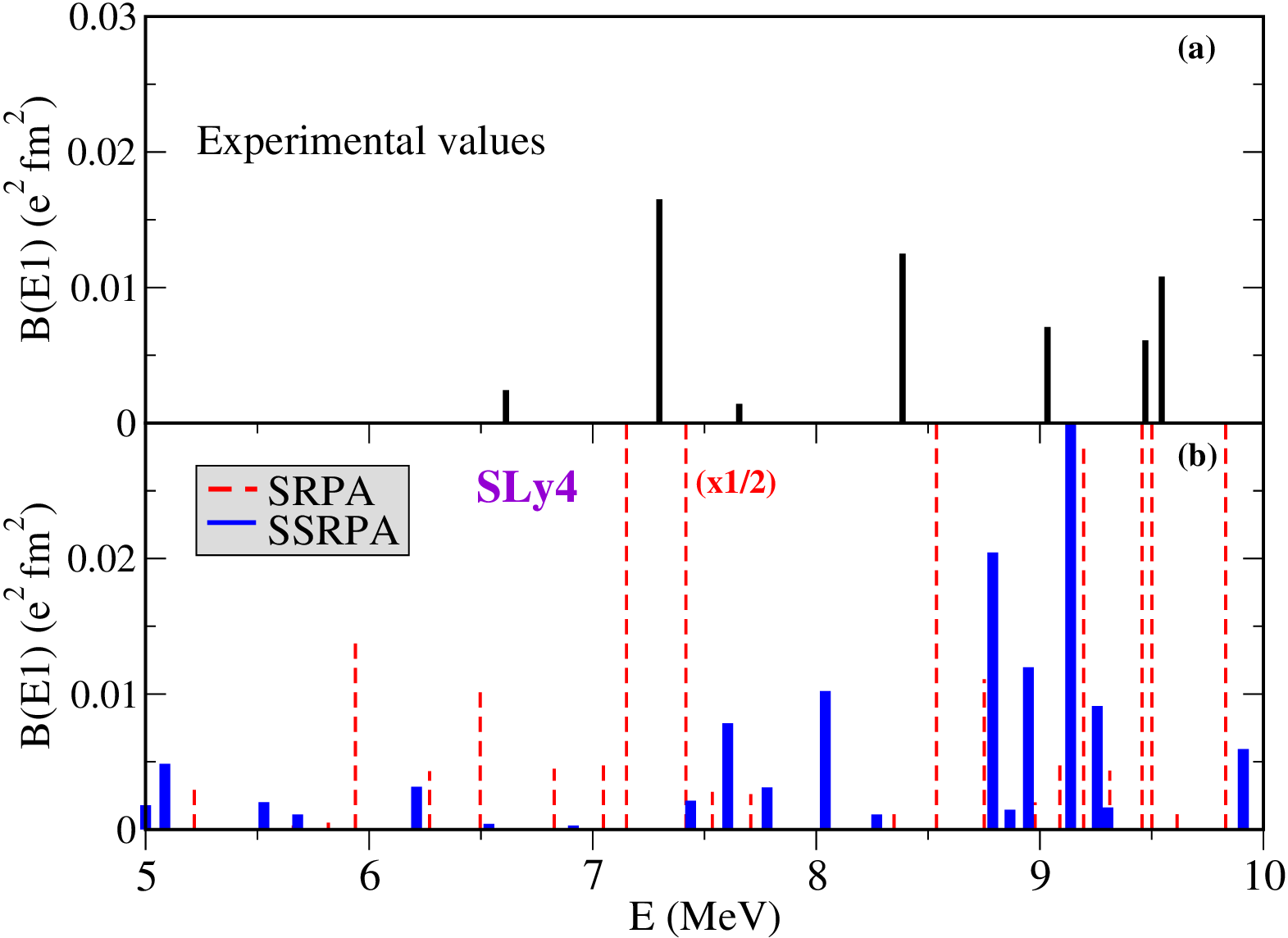}
\caption{Same as in Fig. \ref{fig1}, but the theoretical results are obtained this time with the Skyrme parametrization SLy4.}
\label{fig2}
\end{figure}
A cutoff of 100 MeV is used in building the 1p1h configurations,
ensuring a full presevation of the isoscalar and isovector Energy Weighted Sum Rules (EWSRs).
Deviations of less than 1\% are found in both cases in RPA. 
Owing to the fully self--consistent procedure that we adopt here, possible contaminations of the low--lying spectrum with spurious
components, that might be present for instance in the results discussed in Ref. \cite{gambapygmy}, are now removed. 
Within the SRPA model, the EWSRs are satisfied as in RPA, with deviations less than 1\%. 
On the other side, the SSRPA model provides by construction the same inverse moment $m_{-1}$ as in RPA, but not the same moment $m_1$ and deviations of $\approx 5-7$ \% are typically found for the EWSRs. 
In spite of the fully self--consistency, the SRPA model leads to a spurious state with imaginary energy. This is related to the stability problem investigated in Ref. \cite{papa}. Using the subtraction procedure, such a stability problem is cured and the Thouless theorem is satisfied \cite{tse2}. No imaginary solutions are found in SSRPA calculations. 

We use a cutoff of 60 (SGII) and 70 (SLy4) MeV in the 
2p2h configurations for both the construction of the matrix to be 
diagonalized and the evaluation of the corrective term induced by the 
subtraction procedure \cite{gamba2015}. The two different cutoffs are chosen to 
 guarantee in the two cases stable results both in the
low-lying energy spectrum and in the GDR region. Starting from these two cutoff values, 
SSRPA results are practically cutoff independent, as already discussed in 
 Ref. 
\cite{gamba2015}, 
because the subtractive term removes the ultraviolet diverging contribution. 
Owing to the huge number of 2p2h configurations ($\sim 10^7$), we employ a diagonal approximation
in the subtraction procedure (the diagonal approximation is adopted in the 2p2h sector of the matrix used in the subtractive term, but not in the matrix to be diagonalized). This approximation was tested in  Ref. \cite{gamba2015}
where we showed results very close to those obtained within the exact subtraction procedure. 
Different is the case where the diagonal approximation is employed also in the matrix to be diagonalized leading to significantly different results and indicating that the contribution of the residual interaction among the 
2p2h configurations is crucial there to describe the fragmentation of the low-lying
response and the centroid energy in the  GDR region.

We start by analyzing the low--energy part of the spectrum, between 5 and 10 MeV. 
We show in Fig. \ref{fig1}(a) the experimental transition probabilities 
$B(E1)$ \cite{low}. The corresponding theoretical results are displayed in Fig. 
\ref{fig1}(b). One observes that the total strength 
provided by the standard SRPA is much higher than the experimental one as 
already found and discussed in Ref. \cite{gambapygmy}. We stress that,
in the RPA case, with both the interactions used here, there is no strength  
below 10 MeV.  
The theoretical prediction given by the SSRPA model, on the other side, is in a 
very satisfactory agreement with respect to the experimental data. The 
fragmentation of the states follows very well the experimental distribution in 
the energy position of the main peaks. The total strength integrated between 5 
and 10 MeV is shown in Table I. The first column shows the experimental result 
and the second and third columns the corresponding values obtained with the SRPA 
and the SSRPA models, respectively, employing the Skyrme parametrization SGII. 
Whereas the SRPA value is eight times larger than the experimental result, 
the SSRPA summed $B(E1)$ is very close to the measured value. A 
similar behavior is found also for the EWSR
integrated up to 10 MeV, shown in the second line of Table I.  

\begin {table} 
\begin{center}
\begin{tabular}{cccccc}
                     \hline
\hline
                     & Exp           & SRPA & SSRPA & SRPA & SSRPA\\
  &   &     SGII         &  SGII &   SLy4 & SLy4  \\
\hline
$\sum B(E1)$         & 0.068       & 0.563 & 0.078 & 1.012& 0.126\\
                     & $\pm$ 0.008 &       &       &      &       \\
\hline
$\sum_i E_i B_i(E1)$ & 0.570       & 4.618& 0.621 & 8.795& 1.062\\
                     & $\pm$ 0.062 &       &       &      &       \\
\hline
\hline
\end{tabular}
\end{center}
\caption{ Experimental and theoretical $\sum B(E1)$ in ( e$^{2}$ fm$^{2}$) and $\sum_i E_i B_i(E1)$ in (MeV e$^{2}$ fm$^{2}$) 
summed between 5 and 10 MeV.}
\end {table} 

Figure \ref{fig2} illustrates the same results as Fig. \ref{fig1} but for the SLy4 parametrization. One observes also in this case a clear improvement of the results when the subtraction procedure is applied since this procedure induces a considerable reduction of the strength. With this Skyrme parametrization, results are however less satisfactory than those obtained with the parametrization SGII in the comparison with the experimental results. 
The fragmentation of the strength is less well reproduced and the strength summed between 5 and 10 MeV is less close to the experimental value (Table I) compared to the SGII case.
A clear improvement produced by the subtraction is anyway observed: the summed strength between 5 and 10 MeV without any subtraction is 15 times larger than the experimental value (fourth column of Table I) and is reduced to twice the experimental measurement by the subtraction procedure (fifth column of Table I).

\begin{figure}
\includegraphics[width=0.4\paperwidth]{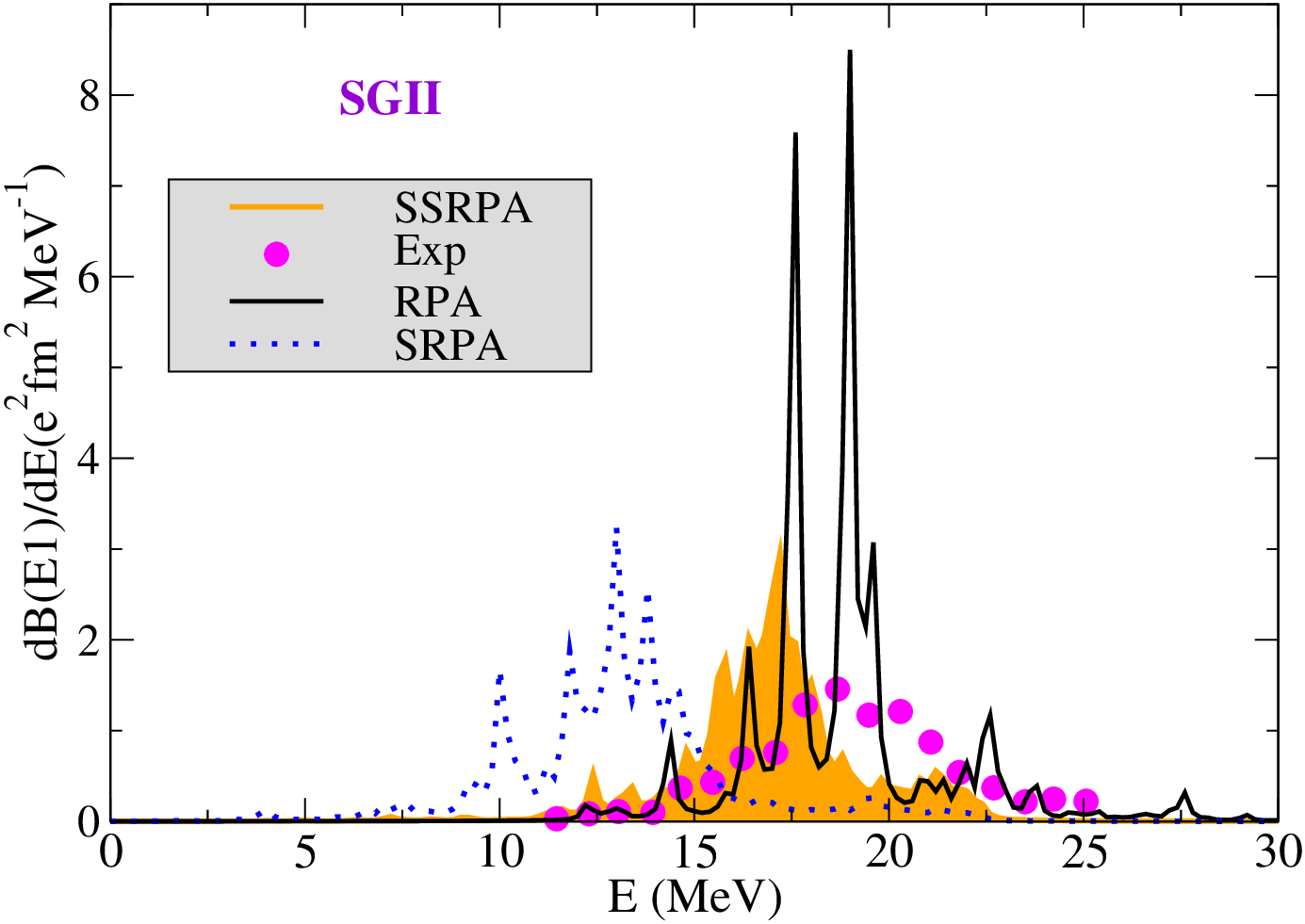}

\caption{Dipole strength distributions evaluated with RPA (solid black line), SRPA (blue dotted line), and SSRPA (orange line and area), compared with the experimental distributions (magenta circles) of Ref. \cite{osaka}. The SGII Skyrme interaction is used. }
\label{fig3}
\end{figure}

We now move to the region between 15 and 25 MeV, where the GDR is located. 
Figure \ref{fig3} shows the strength distributions calculated with the parametrization SGII in RPA, SRPA, and SSRPA, compared with the experimental distributions of Ref. \cite{osaka}. To help visualizing the theoretical distributions, a folding with a Lorentzian having a width of 0.25 MeV is performed. This folding produces some artificial spreading in the RPA case. For both SRPA and SSRPA, a more pronounced spreading is found, which is this time not artificial because it corresponds to a physical width  produced by the extremely dense distribution generated by the 2p2h configurations. It can be observed that the width is indeed in a satisfactory agreement with the experimental distribution. Reference \cite{osaka} 
reports the experimental centroid energy $E_C$ and width $\Gamma$, which are equal to 18.9 $\pm$ 0.2 and 3.9 $\pm$ 0.4 MeV, respectively.  
We have computed the theoretical centroid energies and widths using the expressions
\begin{equation}
\nonumber
E_C = \frac{m_1}{m_0}, \;\;\; \Gamma_C=\sqrt{m_2 / m_0 -(m_1/m_0)^2},
\end{equation}
where $m_k$ represents the moment of order $k$ integrated in the energy region of interest. In the case of SRPA--based calculations, where the strength is more fragmented, we have compared these centroids and widths with those extracted by fitting our distributions with a Lorentzian. No significant differences were found between the results provided by the two methods. 

\begin{figure}
\includegraphics[width=0.4\paperwidth]{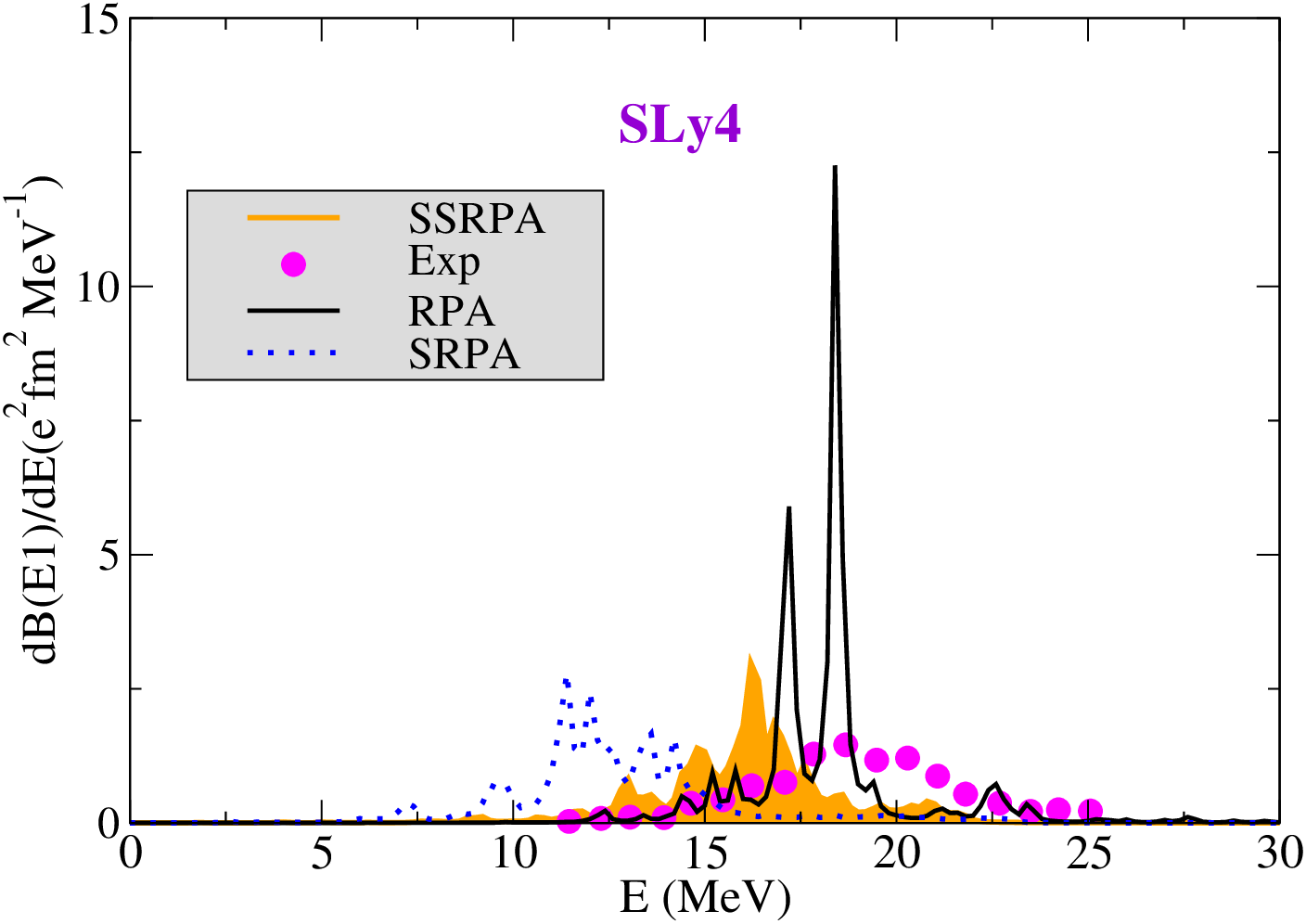}

\caption{Same as in Fig. \ref{fig3} but with the Skyrme interaction SLy4.}
\label{fig4}
\end{figure}

RPA calculations, which describe reasonably well the centroid of the strength distribution ($E_C$ = 18.6 MeV), do not provide any broad distribution. One observes that the distribution is shared between two main discrete states. The SRPA model has the advantage of leading in a natural way to a physical fragmentation ($\Gamma=$ 2.4 MeV), but the spectrum is, as expected, too much shifted by several MeV to lower energies with respect to RPA and to the experimental energies ($E_C$ = 13.5 MeV). The SSRPA model corrects in a significant way this shift ($E_C=$ 17.4 MeV), although a slight underestimation of $\sim$ 1.5 MeV with respect to the experimental centroid is still present. The width remains more or less unchanged with respect to the SRPA case ($\Gamma$ = 2.5 MeV). The missing width with respect to the experimental value ($\sim$ 1 MeV) 
could be generated by the coupling to the continuum (escape width),  which is not taken into account in our case,
or by the coupling with more complex configurations (3p3h, 4p4h, ...). 

Figure \ref{fig4} describes the same quantities as in Fig. \ref{fig3} but in this case the theoretical predictions are obtained with the parametrization SLy4. Similar remarks may be done as for the previous case. The theoretical centroid energies are 
18.0, 13.4, and 16.2 MeV with RPA, SRPA, and SSRPA, respectively. The widths are equal to 2.5 and 2.1 MeV with SRPA and SSRPA, respectively. Also for the GDR part of the spectrum we observe that the parametrization SLy4 provides less satisfactory results than the Skyrme interaction SGII. In particular, the SSRPA centroid energy is now more than 2.5 MeV lower 
than the experimental centroid.   

\begin{figure}
\includegraphics[width=0.4\paperwidth]{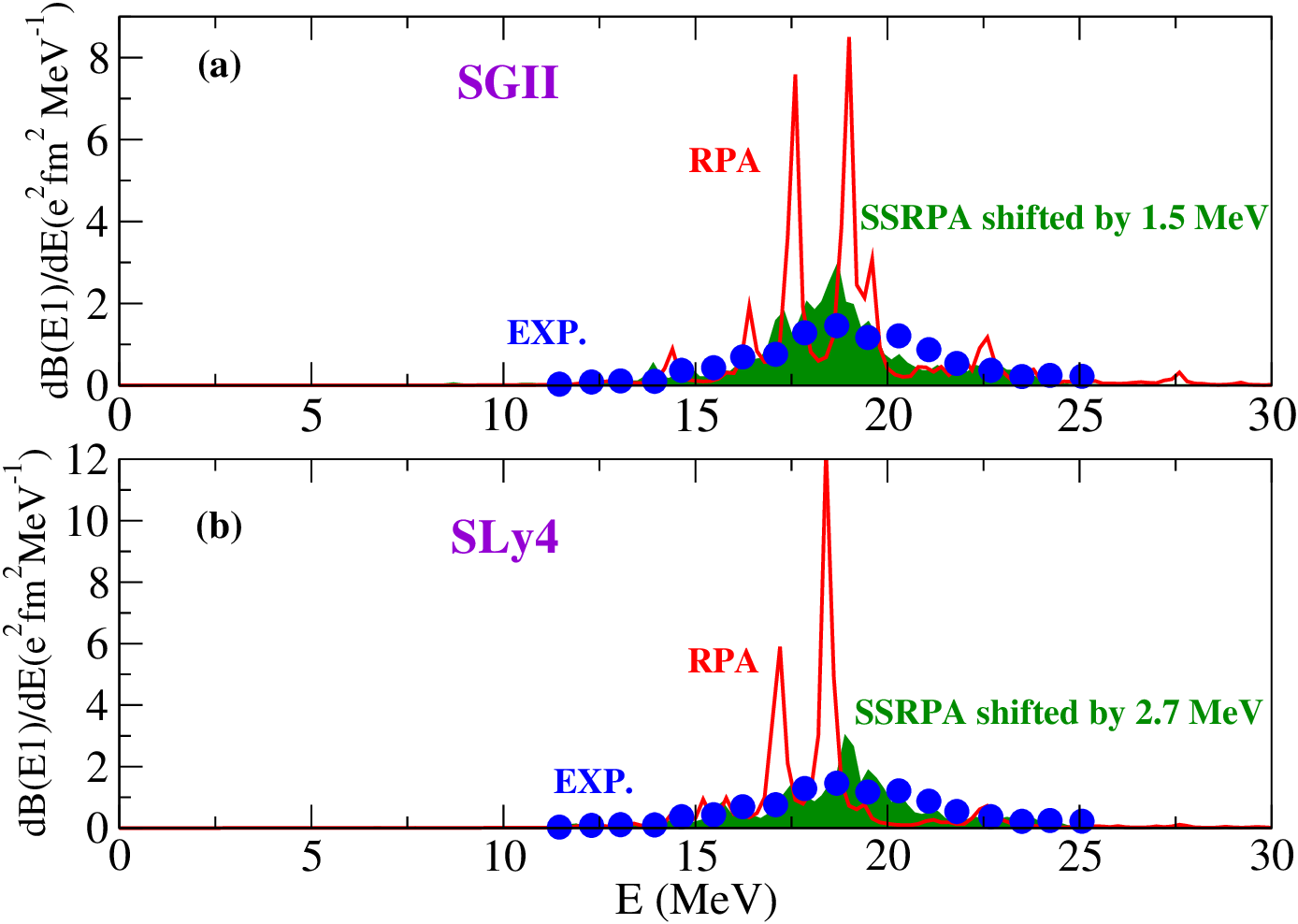}
\caption{(a) SSRPA results shifted by 1.5 MeV (green area) compared with the RPA strength (red line) and with the experimental values (blue circles), obtained with the parametrization SGII. (b) Same as in (a) but the results are obtained with the parametrization SLy4 and the shift of the SSRPA spectrum is larger (2.7 MeV).  }
\label{fig5-1}
\end{figure}

Figure \ref{fig5-1} displays in the two panels the SSRPA strength distributions obtained with SGII (a) and SLy4 (b), shifted upwards respectively by 1.5 (a) and 2.7 (b) MeV, together with the RPA strength distributions and the experimental data. This plot clearly illustrates how well the width of the resonance is described and shows the significant improvement with respect to RPA.

Using now the theoretical $B(E1)$ probabilities, we may calculate the electric dipole polarizability
\begin{equation}
\alpha_D = \frac{8 \pi}{9} \int \frac{B(E1,E_x)}{E_x} dE_x.
\label{polari}
\end{equation}

\begin{figure}
\includegraphics[width=0.4\paperwidth]{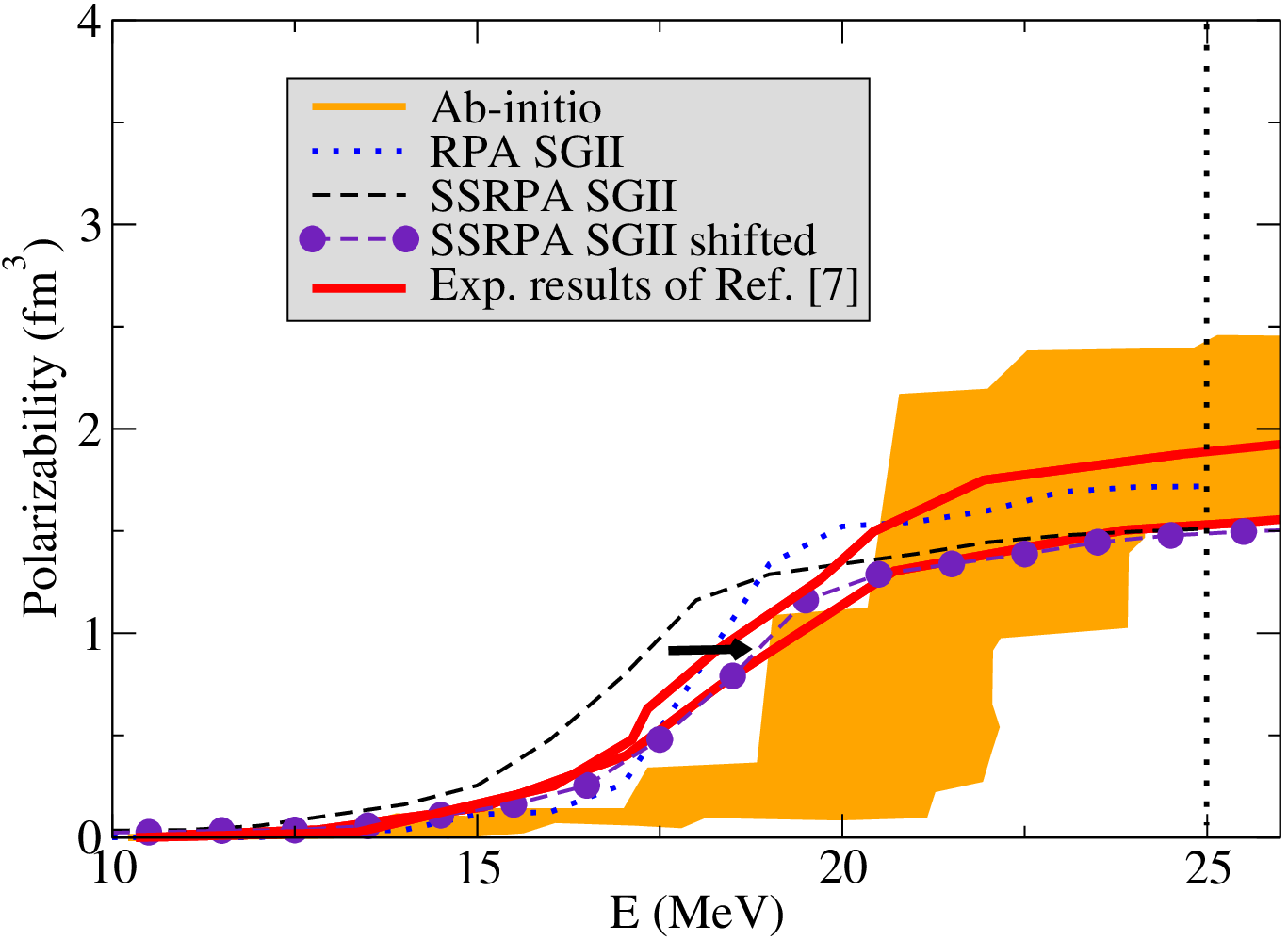}

\caption{Electric dipole polarizability as a function of the excitation energy. The orange area corresponds to {\it ab--initio} results and the area between the two solid red lines to experimental results reported in Ref. \cite{osaka}. In particular, the measurement illustrated in Ref. \cite{osaka} provides the values in the band up to the vertical dashed line located at 25 MeV. Results obtained with the RPA (blue dotted line) and the SSRPA (black dashed line) models are displayed. The violet circles and dashed line represent these latter results shifted upwards by 1.5 MeV. The Skyrme interaction SGII is used.}
\label{fig5}
\end{figure}

The photoabsorption cross section reported in Ref. \cite{osaka} is measured in the range from 10 to 25 MeV. The reported contribution to the electric dipole polarizability between 10 and 25 MeV is 1.73 $\pm$ 0.18 fm$^3$. The contribution found experimentally below 10 MeV is negligible. With the parametrization SGII (SLy4) we obtain, below 10 MeV, $\alpha_D = $ 6 $\cdot$ 10$^{-4}$ and 3 $ \cdot$ 10$^{-3}$ (10$^{-3}$ and 5 $\cdot 10^{-2}$)
fm$^3$ with RPA and SSRPA, respectively.
These values are indeed negligible, as for the experimental case. 
 We display in Fig. \ref{fig5} the values of $\alpha_D$ calculated by varying the upper limit of the integral in Eq. (\ref{polari}) up to 25 MeV. 
We show only our result obtained with the parametrization SGII because, as already noticed, this represents our most satisfactory prediction.  
The obtained curve is compared with the corresponding experimental and theoretical results extracted from Fig. 4(b) of Ref. \cite{osaka}. Integrating up to 
25 MeV is not enough to obtain a saturated value for the electric dipole polarizability and this is the reason why the curves in Fig. 4(b) of Ref. \cite{osaka} are displayed up to 60 MeV, in such a way that a converged value can be deduced for the polarizability. 
To do this for the experimental values, the authors have combined their experimental results (up to 25 MeV) with results deduced from data available for $^{40}$Ca \cite{arhens}, using a procedure inspired by Ref. \cite{hashi}: they have taken the $^{40}$Ca photoabsorption data of Ref. \cite{arhens} and shifted them by the difference of the centroids expected according to the formula 
$E_C=31.2 A^{-1/3}+20.6 A^{-1/6}$ \cite{mass}. 
In our case, 
we are however limited by the huge numerical effort required by SRPA--based calculations. We perform then our integration up to 25 MeV and we compare our results only to those that correspond to the measurement of Ref. \cite{osaka} (area between the red solid lines up to 25 MeV, indicated by the vertical dotted line). 

The theoretical results reported in Fig. 4(b) of Ref. \cite{osaka} are obtained 
with {\it ab--initio} coupled--cluster calculations \cite{cc1,cc2}. They are 
displayed in Fig. \ref{fig5} as an orange band, which corresponds to the use of 
different chiral Hamiltonians with the interactions of Ref. \cite{hagen} that 
lead to reasonable saturation properties for symmetric nuclear matter.  
Comparing the {\it ab--initio} curves with the experimental band (area between 
the two dashed lines) we easily understand that the {\it ab--initio} strength 
distributions, peaked around 20 MeV, are much less spread around their centroid 
energies. In the theoretical band, the dipole polarizability increases indeed 
very rapidly around 20 MeV whereas, experimentally, $\alpha_D$ increases more 
smoothly in the energy region where the giant resonance is spread. Moreover, it 
is clear from the figure that the ab-initio theoretical calculations 
overestimate the centroid energy which is located at more than $\sim$ 20 MeV. 
One observes that our Skyrme SGII SSRPA curve follows much better the 
experimental profile in the energy region where the GDR is spread: the fact that 
the slope is well reproduced indicates that the spreading is well described. The 
 underestimation of the centroid is visible in a global small shift of the 
curve with respect to the experimental band. The violet circles and dashed line represent 
the same curve shifted upwards by 1.5 MeV to better illustrate how well the slope 
is reproduced. 
The RPA results have the same drawback as the {\it ab--initio} results showing a 
stiff increase of the polarizability in a narrower region of energy. We observe 
that the Skyrme RPA value for the polarizability at 25 MeV is located just in the middle 
of the experimental band, indicating a good agreement with respect to the 
measurement. It is worth noticing that, integrating up to higher values of the 
energy, where the polarizability reaches its converged value, the RPA and SSRPA 
polarizabilities must converge to the same value by construction: the 
subtraction procedure is indeed based on the requirement that the moment 
$m_{-1}$ is the same in RPA and in SSRPA. Of course, it is expected a slower 
convergence for the SSRPA values. A small difference between the two 
calculations is expected, due to the adopted diagonal approximation in the 
corrective term of the subtraction procedure (see upper panel of figure 13 of Ref. \cite{gamba2015}. 

\section{Conclusions}
To summarize, we applied for the first time the Skyrme SSRPA model in a fully
self-consistent scheme, that is, including all the terms of the residual interaction,
allowing thus for a reliable description of the dipole response. We analysed both the low-lying
part of the spectrum as well as the GDR region. In the first case, RPA results do not provide
any strength below 10 MeV. The inclusion of the 2p2h configurations within the SRPA scheme 
allows for a better description. However, the total strength
between 5 and 10 MeV is strongly overestimated. We showed that, by using the subtraction
procedure recently implemented in the SSRPA scheme, a very satisfactory agreement
with respect to the experimental results is found, both concerning the total strength and the 
corresponding EWSR. We also showed that the SSRPA provides a good description of the GDR properties, in particular
regarding the width that cannot be described by the RPA approach. Finally,
the consistent description of the dipole response both at low and higher energy,
allows for a more satisfactory description of the dipole polarizability compared
to the RPA and {\it ab--initio} calculations. This improvement over the other theoretical approaches consists in a more reliable description of the slope of the electric polarizability, displayed as a function of the excitation energy, in the region 
around the centroid of the GDR, where the excitation is spread. 





\end{document}